\documentclass[a4paper,10pt]{article}
\usepackage{graphicx}
\usepackage[small]{caption}
\title{Recursive Equations for Majorana Currents}
\author{R. Kleiss, G. van den Oord}
\begin{document}

\maketitle

\begin{abstract}
A recursive computation of scattering amplitudes including Majorana fermions requires a consistent definition of the fermion flow, which is introduced by Denner et al. in a diagrammatic setting. A systematic treatment in the off-shell current formalism is proposed, which involves explicit reversal of fermion currents.
\end{abstract}
\begin{figure}
 \includegraphics[width=\textwidth]{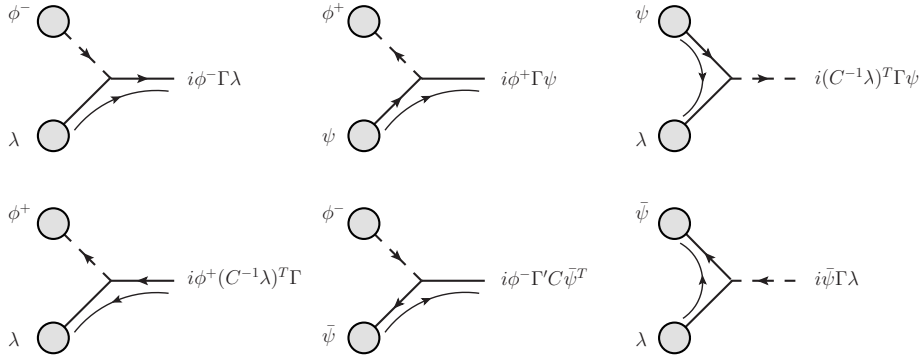}
\caption[small]{Majorana-Dirac vertex terms in the recursion relations. The momentum flows from the left to the right.}
\label{vertices}
\end{figure}

\vspace{2 cm}
Recursive algorithms \cite{BerendsNPB306,CaravagliosPLB358} have proved themselves superior to diagrammatic methods in automatized numerical tree-level calculations of matrix elements. Especially for high-multiplicity processes, an iterative computation of off-shell currents is feasible, as its complexity only grows exponentially, whereas the number of Feynman diagrams grows factorial. To extend these calculations to supersymmetric theories, an efficient and consistent treatment of Majorana particles is necessary. The difficulty here is to determine whether the Majorana current should behave as a particle or an antiparticle. In \cite{DennerPLB291} it was shown that an assignment of a continuous fermion flow in each diagram solves this problem unambiguously. This assignment is not trivial to generalize to off-shell currents because they are typically used by several Feynman diagrams that may combine the fermion currents in different ways. It is therefore appropriate to choose a particular flow direction for the currents, and in the combination procedure 'reverse' the current when needed. This reversal involves an explicit multiplication with the charge conjugation matrix, which fixes the relative phase of spinor-anti-spinor wave functions:
\begin{equation}\label{ccwf}
 v(p,s)=C\bar{u}^T(p,s)\,,\qquad u(p,s)=C\bar{v}^T(p,s)\,.
\end{equation}
These operations can be generalized to off-shell fermionic currents in a straightforward manner. A generic column-type internal spinor
\begin{equation}
 \psi=V^{i_1\ldots i_n}S(p_1)\Gamma_{i_1}\ldots S(p_n)\Gamma_{i_n} u(p,s)
\end{equation}
 will upon reversal yield the row spinor
 \begin{equation}
  (C^{-1}\psi)^T=V^{i_1\ldots i_n}\bar{v}(p,s)\Gamma_{i_n}'S(-p_n)\ldots\Gamma'_{i_1}S(-p_1)\,.
 \end{equation}
where $\Gamma'_i$ denote the reversed Dirac matrices (see e.g. \cite{DennerPLB291}). 
Similarly an outgoing anti-fermion current is reversed by interchanging $u$ and $v$ in the above expressions, and row-spinor type current $\bar{\psi}$ transforms to a column spinor by taking $C\bar{\psi}^T$. To minimize the number of reversals in the recursion, it is natural to choose the flow direction for \emph{Dirac} fermions parallel to the fermion number flow. Moreover, one can choose the Majorana particles in the algorithm to exhibit the same behavior as Dirac fermions.
\begin{figure}[ht]
 \includegraphics[width=\textwidth]{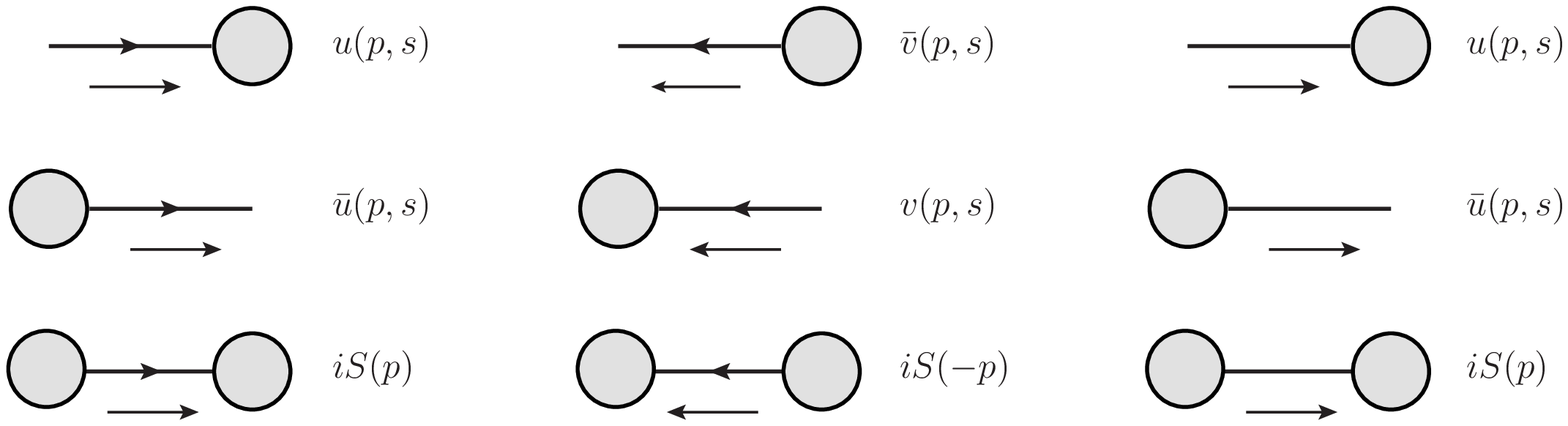}
\caption[small]{Spinor wave functions and propagators in a recursion involving Dirac and Majorana fermions. The momentum runs from left to right.}
\label{spinors&props}
\end{figure}
 These rules determine unambiguously all spinor wave functions and propagators in the procedure, as is shown in fig. \ref{spinors&props}. They also constrain the combination rules for the off-shell currents; to ensure the fermion flow continuity, every combination must result in either a Dirac or Majorana fermion with flow direction parallel to the momentum, or a Dirac anti-fermion with the flow opposite to the momentum. This involves a reversal of the current according to eq. (\ref{ccwf}) whenever a Majorana fermion combines with a boson to a Dirac anti-fermion or vice versa an anti-fermion fuses to a Majorana particle. When two fermion currents combine to a scalar, we choose to reverse the Majorana current whenever the flow is not continuous, i.e. the case where a Dirac and a Majorana fermion current combine (cf. fig. \ref{vertices}).
It should be noted that for outgoing external particles, the vertex rules contain an extra reversal of the spinor wave function.

\begin{figure}[h]
\centering
 \includegraphics[width=8 cm]{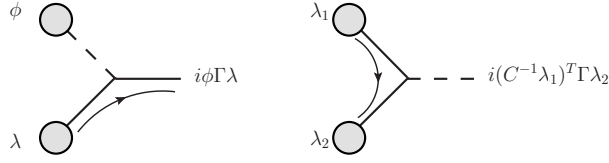}
\caption[small]{Majorana-Majorana vertex terms in the recursion relations, with the choice to reverse the first current when creating a bosonic current.}
\label{MMvertex}
\end{figure}

The recursion relations concerning Majorana-Majorana-boson vertices are not unambiguously fixed by the above rules. When combining 2 Majorana currents to a boson, it remains arbitrary which one is reversed, since only couplings that are symmetric under reversal can appear in the Lagrangian. Figure (\ref{MMvertex}) shows the recursive relations in the case one chooses to reverse the first current.

\begin{figure}[ht]
 \centering
\includegraphics[width=4.5 cm]{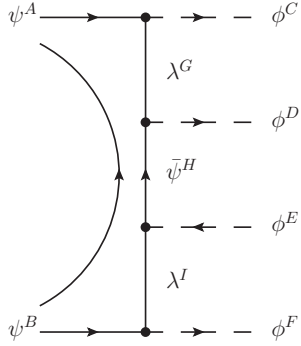}
\caption[small]{Feynman diagram contributing to $\psi\psi\rightarrow\phi\phi\phi\phi$.}
\label{2fto4s}
\end{figure}

As an example, consider the fermion-number-violating process $\psi\psi\rightarrow 4$ scalars. The contribution of the Feynman diagram () can be easily written down using the compact Feynman rules as
\begin{equation}\label{Feynform}
 i\mathcal{M}=i^7h^i_{GAC}h^j_{GHD}h^k_{IHE}h^{\ell}_{IBF}\bar{v}_A\Gamma'_iS(-P)\Gamma_jS(-Q)\Gamma_kS(-R)\Gamma_{\ell}u_B
\end{equation}
where $P=p_A-p_C$, $Q=P-p_D$ and $R=Q-p_E$. With a recursive method, the computation starts by constructing the fermion wave function $u_A$. Then the first internal current is computed, which yields
\begin{equation}
 \lambda_G=i^2h^i_{GAC}S(P)\Gamma_iu_A\,.
\end{equation}
 The combination of this current with the second scalar to a Dirac anti-fermion involves the reversal of $\lambda_G$,
\begin{eqnarray}
 \bar{\psi}_H &=& i^2h^j_{GHD}(C^{-1}\lambda_G)^T\Gamma_jS(-Q)\\ \nonumber
&=& i^4 h^i_{GAC}h^j_{GHD}\bar{v}_A\Gamma'_iS(-P)\Gamma_jS(-Q)
\end{eqnarray}
The production of the second internal Majorana current involves a charge conjugation once more,
\begin{eqnarray}
 \lambda_I &=& i^2h^k_{IHE}S(R)\Gamma'_kC\bar{\psi}_H^T\\ \nonumber
&=& i^6 h^i_{GAC}h^j_{GHD}h^k_{IHE}S(R)\Gamma'_kS(Q)\Gamma'_jS(P)\Gamma_iu_A 
\end{eqnarray}
where we have used that $(\Gamma_i')'=\Gamma_i$. One quickly find the expression (\ref{Feynform}) upon reversing this current and contracting with $ih^{\ell}_{IBF}\Gamma_{\ell}u_B$.

In conclusion, we have proposed a method to include Majorana particles in recursive computations of scattering amplitudes. The algorithm involves explicit reversal of fermionic off-shell currents. Automatizing this procedure will introduce extra operations on the currents, which is unavoidable since the fermion current is defined on a diagrammatic basis. The overhead of the charge conjugation can be minimized by absorbing it in the vertex contraction routine.

\end{document}